\documentclass[twocolumn]{aastex61}

\usepackage{epstopdf}

\newcommand{\msun}{\rm{M}_\odot}
\newcommand{\mste}{\textit{M}_{\star}}

\newcommand{\gala}{\texttt{GALAPAGOS }}
\newcommand{\galfit}{\textit{galfit }}
\newcommand{\lcdm}{$\mathrm{\Lambda C DM\,}$}

\newcommand{\gin}{$\mathrm{\gamma_{in\,}}$}

\newcommand{\mlow}{$\textit{M}_{\star}^{low}$}
\newcommand{\mhi}{$\textit{M}_{\star}^{high}$}

\begin{document}

\title{Mass profile decomposition of the Frontier Fields cluster MACS J0416$-$2403. Insights on the dark-matter inner profile. }

\author{Annunziatella M.}

\affiliation{INAF-Osservatorio Astronomico di Trieste, via G. B. Tiepolo 11}

\author{Bonamigo M.}
\affiliation{Dark Cosmology Centre, Niels Bohr Institute, University of Copenhagen, Juliane Maries Vej 30, 2100 Copenhagen, Denmark}

\author{Grillo C.}
\affiliation{Dipartimento di Fisica, Universit\`a degli Studi di Milano, via Celoria 16, I-20133 Milan, Italy}
\affiliation{Dark Cosmology Centre, Niels Bohr Institute, University of Copenhagen, Juliane Maries Vej 30, 2100 Copenhagen, Denmark}

\author{Mercurio A.}
\affiliation{INAF-Osservatorio Astronomico di Capodimonte, Via Moiariello 16 I-80131 Napoli, Italy}

\author{Rosati P.}
\affiliation{Dipartimento di Fisica e Scienze della Terra, Univ. degli Studi di Ferrara, via Saragat 1, I-44122, Ferrara, Italy}

\author{Caminha G.}
\affiliation{Dipartimento di Fisica e Scienze della Terra, Univ. degli Studi di Ferrara, via Saragat 1, I-44122, Ferrara, Italy}
\affiliation{INAF - Osservatorio Astronomico di Bologna, via Gobetti 93/3, 40129 Bologna, Italy}

\author{Biviano A.}
\affiliation{INAF-Osservatorio Astronomico di Trieste, via G. B. Tiepolo 11}

\author{Girardi M.}
\affiliation{Dipartimento di Fisica, Univ. degli Studi di Trieste, via Tiepolo 11, I-34143 Trieste, Italy}

\author{Gobat R.}
\affiliation{Korea Institute for Advanced Study, KIAS, 85 Hoegiro, Dongdaemun-gu Seoul 130-722, Republic of Korea}

\author{Lombardi M.}
\affiliation{Dipartimento di Fisica e Scienze della Terra, Univ. degli Studi di Ferrara, via Saragat 1, I-44122, Ferrara, Italy}

\author{Munari E.}
\affiliation{INAF-Osservatorio Astronomico di Trieste, via G. B. Tiepolo 11}

\correspondingauthor{Annunziatella}
\email{marianna.annunziatella@gmail.com}

\begin{abstract}
We present a high resolution dissection of the two-dimensional total mass distribution in the core of the Hubble Frontier Fields galaxy cluster MACS J0416.1$-$2403, at $z \mathrm{\, =\, 0.396}$. 
We exploit  HST/WFC3 near-IR (F160W) imaging, VLT/MUSE spectroscopy, and Chandra data to separate the stellar, hot gas, and dark-matter mass components in the inner 300 kpc of the cluster. We combine the recent results of our refined strong lensing analysis, which includes the contribution of the intracluster gas, with the modeling of the surface brightness and stellar mass distributions of 193 cluster members, of which 144 are spectroscopically confirmed. 
We find that moving from 10 to 300 kpc from the cluster center the stellar to total mass fraction decreases from 12\% to 1\% and the hot gas to total mass fraction increases from 3\% to 9\%, resulting in a baryon fraction of approximatively 10\% at the outermost radius. 
We measure that the stellar component represents $\sim$ 30\%, near the cluster center, and 15\%, at larger clustercentric distances, of the total mass in the cluster substructures. 
We subtract the baryonic mass component from the total mass distribution and conclude that within 30 kpc ($\sim$ 3 times the effective radius of the BCG) from the cluster center the surface mass density profile of the total mass and global (cluster plus substructures) dark-matter are steeper and that of the diffuse (cluster) dark-matter is shallower than a NFW profile. Our current analysis does not point to a significant offset between the cluster stellar and dark-matter components. 
This detailed and robust reconstruction of the inner dark matter distribution in a larger sample of galaxy clusters will set a new benchmark for different structure formation scenarios. 
\end{abstract}

\section{Introduction}
\label{s:intro}

One of the main achievements of the current $\mathrm{\Lambda C DM}$ cosmological paradigm is to be 
able to describe the large-scale distribution of matter in the Universe at different epochs 
\citep{springel2006}. 
Cosmological N-body simulations implemented within the \lcdm  paradigm have provided precise 
predictions on the formation and evolution of dark-matter halos over a wide range of scales. A key 
result of these simulations is that  dark-matter halos of all masses have ``universal'' mass 
density profiles  that are well described by a simple law with a central cusp $\rho (r) \sim 
r^{-1}$, and a steeper slope, $\rho (r) \sim r^{-3}$, at large radii \citep[the so 
called NFW profile;][]{navarro1996}.
Despite the great success of the \lcdm  predictions, some discrepancies with available 
observations still exist. \\
Some tension between the observed and predicted values of the inner slope of the 
dark-matter mass density profile has been detected at two extremes of the halo mass distribution: 
dwarf galaxies and galaxy clusters. 
In the past few years, significant progress has been made towards the measurement of the value of the inner logarithmic slope ($\mathrm{\gamma_{in}}$) of the dark-matter mass profile in clusters, but, in some cases, the results obtained 
by different groups on same clusters are still controversial. 
For example, \citet{okabe2016} find that 50 X-ray luminous galaxy clusters with good gravitational lensing  data have a stacked total mass density profile consistent with the NFW profile from the inner 
core to the virial radius. Similarly, \cite{umetsu2016} conclude that the stacked  total  
mass density profile of 20 massive clusters in the Cluster Lensing And Supernova survey with Hubbe 
\citep[CLASH; ][]{postman2012} survey is well described by a NFW profile. 
On the other hand, 
\cite{newmana,newmanb} find that the total mass density profile in the center of clusters closely 
follows the NFW profile but, once the contribution of the stellar component is 
subtracted, the inferred dark-matter mass density profile is significantly flatter than a NFW profile.
On smaller scales, dwarf galaxies are studied in the same context because their very high 
mass-to-light ratios suggest that baryonic effects may have been minor in their mass assembly 
history. 
Dynamical analyses of dwarf and low surface brightness galaxies seem to favor massive dark matter 
haloes with surprisingly shallow or cored ( \gin $\ll$ 1) inner density profiles 
\citep[e.g., ][]{amorisco2012, agnello2012}, whereas much steeper (\gin $\sim$ 2) profiles are preferred in 
massive early-type galaxies from strong gravitational lensing and stellar population modeling 
\citep[e.g. ][]{grillo2012}. \\
These debated results are also known as the dark-matter cusp-core problem.
The value of \gin could contain important information about the nature of the dark matter. For 
example, if the dark-matter particles were self-interacting rather than effectively collisionless, 
with a sufficiently large self-interaction cross-section, the inner halo mass density profile 
should be shallower than a NFW profile \citep{navarro1996}, in the absence of 
baryonic effects \citep{rocha2013}. 
A major leap forward in addressing these fundamental questions can only be made by obtaining 
homogeneous, high quality data on a sizable and unbiased sample of astrophysical objects. 
Clusters of galaxies, by virtue of their position at the high end of the mass function, serve 
as giant physics laboratories to explore the role and nature of dark matter, providing unique tests 
of any viable cosmology and structure formation scenario and possible modifications of the laws of 
gravity. Furthermore, massive clusters offer this unique opportunity, as a number of observational 
probes of their mass profiles can be used to robustly check the \lcdm predictions on a large 
dynamical range of densities and distances from the cluster centers.\\
 The main goal of this paper is to disentangle the dark-matter distribution in the massive galaxy 
cluster MACS J0416.1$- $2403 (hereafter M0416) and to measure the values of the inner slope  of the cluster dark-matter halo. In this study, we present for the first time an 
accurate determination of the stellar, hot gas, and total projected mass density profiles out to 
300 kpc from the cluster center.
 Hence we are able to separate the baryonic and 
dark-matter components from the cluster total mass distribution. 
We are also able to evaluate the fractions of the different components relative to the total mass 
of the cluster. \\
This paper is organized as follows. In Sect.~\ref{s:data}, we briefly introduce the photometric and spectroscopic data used in this work. In Sect.~\ref{s:m_prof}, we describe how we derive the stellar mass profile of M0416. In Sect.~\ref{s:baryon_to_total}, 
we analyze the distribution of different cluster components. In Sect.~\ref{s:dm_profile}, we focus on the dark-matter component of M0416. Finally, in Sect.~\ref{s:conclusion}, we draw our conclusions. \\
Throughout this paper, we use  $\mathrm{H_0 \, = \, 70\, km\, s^{-1}\, Mpc^{-1} }$,
$\mathrm{\Omega_M \, = \, 0.3, }$ and $\mathrm{\Omega_{\Lambda} \, =
  \, 0.7 }$. At the cluster redshift, the scale is 321 kpc $\mathrm{arcmin^{-1}}$. All the 
magnitudes used in this work are referred to the AB system.

\section{Data Sample}
\label{s:data}

M0416 is  a massive galaxy cluster first detected by \citet{ebeling2001}. 
This cluster has been imaged with HST for a total of 25 orbits using 16 different filters as a 
part of the CLASH survey. M0416 has also been observed with the VIsible 
Multi-Object Spectrograph (VIMOS) at the ESO/VLT, as part of the ESO Large Programme 
 ``Dark Matter Mass Distributions of Hubble Treasury Clusters and the Foundations of 
$\Lambda$CDM Structure Formation Models'' \citep[CLASH-VLT;][]{rosati2014}. CLASH-VLT collected a 
large sample of spectra for galaxies in the field of view of this cluster, leading to the spectroscopic 
confirmation of $\sim$ 800 cluster members and to the discovery of multiply-imaged background 
sources. More details on VIMOS spectroscopic data 
can be found in \cite{balestra2016}. These data have been used to obtain a precise total reconstruction of the cluster via gravitational strong lensing modeling \citep{grillo2015}. \\
M0416 was then selected to be re-observed, as part of the Hubble Frontier Fields (HFF) initiative 
\citep{lotz2016}, in  ACS/optical  (F435W,  F606W, F814W) and  WFC3/IR  (F105W, F125W, F140W, 
F160W) filters, for a total of 140 orbits, reaching a detection limit of $\sim$ 29 mag 
(AB) at $\mathrm{5\sigma}$ for point-sources. These observations of M0416 were completed in September 2014. In all filters, mosaics are available with 30 and 60 mas pixel scale.\\
M0416 was later observed with the Multi Unit Spectroscopic Explorer (MUSE) at the VLT. In this 
work, we exploit MUSE archival data obtained from two different programs which 
covered the North-East (NE) and South-West (SW) regions of the cluster. 
A detailed description of the MUSE data reduction and analysis is given in \cite{carminha2017}. \\
Here, we use the sample of cluster members also considered in 
\citet[][hereafter Bo17]{bonamigo2017}, including 193 
galaxies, 144 with spectroscopic redshifts, and the others selected  based on their 
$N$-dimensional distance, in color space, from the locus of the spectroscopically confirmed member 
galaxies \citep[see][for more details]{grillo2015}.

\subsection{Stellar masses} \label{ss:mass}
In \cite{grillo2015}, the HST photometry available from the CLASH survey was used to determine 
the stellar mass values of a subsample of our catalog of spectroscopic members. In that paper, the 
images of the cluster in the reddest HST bands (from F435W to F160W) were used to perform a fit of the 
spectral energy distributions (SEDs) of these galaxies. The SED fitting was performed using 
composite stellar population (CSP) models, based on \cite{bruzual&charlot2003} templates, with 
solar metallicity and a \cite{salpeter1955} stellar initial mass function (IMF). The star 
formation histories used were parametrized as delayed exponential functions and the presence of dust was 
taken into account following \cite{calzetti2000}. 
For each galaxy, the best-fit ($\mathrm{\mste^{best}}$) and $\mathrm{1\sigma}$ lower (\mlow) and upper limit (\mhi) values of the  stellar mass were measured. 
An example of a SED is shown in Figure 6 of \cite{grillo2015}.  In \cite{annunziatella2014, annunziatella2016} we have shown that we reached an accuracy of 10\% in stellar masses down to $\mathrm{10^9 \msun}$, thanks to the multi-band HST photometry.

\section{Mass profiles} \label{s:m_prof}
In this section, we describe how we derive the two-dimensional stellar mass distribution of the cluster 
members. We use the following approach: we reconstruct the surface brightness distribution of 
all member galaxies in the reddest HST band  ($\mathrm{F_{160}}$),  then, we use the best-fit values of the 
stellar masses of the subsample of galaxies discussed in Sect.~\ref{ss:mass} to derive an average 
stellar mass-to-light ratio ($\mathrm{\mste/L}$) for all cluster members. Cluster members have a  $\mathrm{\mste/L}$  $\sim $ 0.5 in the $F_{160}$ band, without significant variations over the probed stellar mass range. 
This $\mathrm{\mste/L}$ is hence  used to convert the cluster cumulative luminosity profile into a cumulative stellar mass 
profile.

\subsection{Surface brightness profiles}
\label{ss:galfit}

To determine the surface brightness profile of each cluster member, we use an iterative approach 
based on the two software: \galfit\citep{peng2010_gal} and \gala\citep{barden2012}. 
\textit{Galfit} is a code to model the surfaces brightness profile of galaxies, while \gala is a 
set of procedures that use \galfit to reconstruct the surface brightness profile of all extended 
sources detected by SExtractor in a image 
\citep{bertin1996}. In this automatic run of \galfit, we adopt a Sersic profile for each galaxy. 
In the following, we describe briefly our method. 

\begin{itemize}
\item  We run \gala on the HST image of the cluster, in the $F_{160}$ band, with a pixel size of
60 mas/pixel. We choose not to use the image with the highest angular resolution since we are more
interested in the global surface brightness model of the cluster than in the detailed structure of 
single galaxies. The input PSF is derived from real stars in the HST field from images with the  
30 mas pixel-scale. Therefore, we set the input parameter \texttt{PSF\_OVERSAMPLING} to 2.
\item We then use the parameters coming from the first run of \gala as input parameters to perform 
a Sersic fit of just cluster members and very close galaxies which could affect the photometry of 
the members. To do this, we divide our image into large stamps containing 
approximately 10 galaxies each and use \galfit on this
sub-images. We use segmentation maps as \textit{bad pixel masks} to identify the objects to fit in
each stamp. We also fix the value of the sky background to $2 \times 10^{-3}$, which is the mean 
value that we obtain in empty small regions of the image.  
\item Once the values of the model parameters of all galaxies in each stamp are stable, we perform 
a global fit of all the sources identified in the previous step. The result of this global fit 
shows a diffuse component, mainly between the two brightest cluster galaxies (BCGs) of the cluster, 
that can be associated to the intracluster light (ICL). For this reason, we add an extra source 
modeled in input with a Sersic profile with $n\, =\, 1$. In this step, we fix all sources except those of the ICL and the two BCGs. 
\item Finally, we run \galfit again on the global image using the parameters of the ICL and the 
BCGs determined in the previous step. 
\end{itemize}

\begin{figure*}[t]
\centering
 \includegraphics[width=\linewidth]{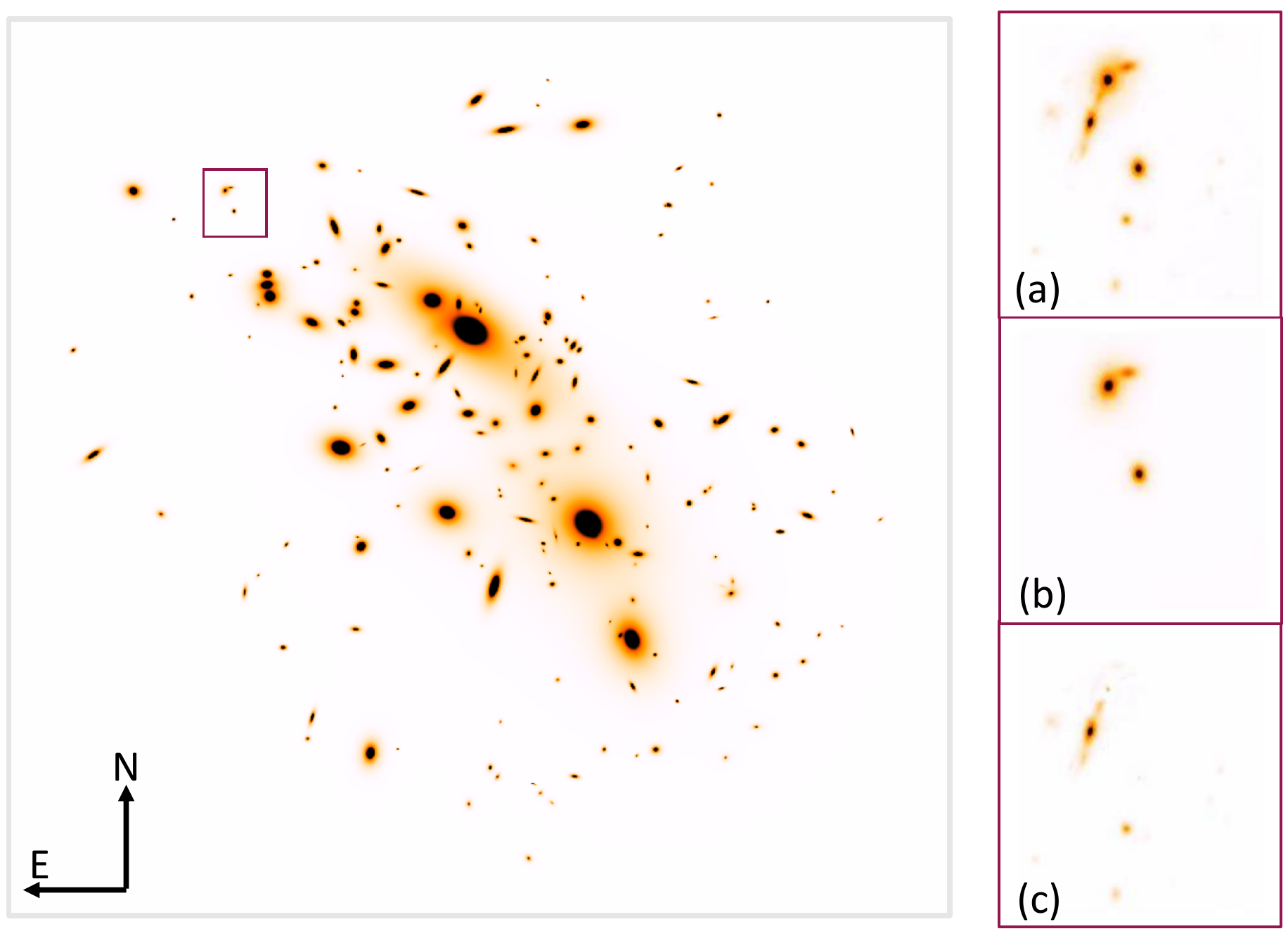}
\caption{Left: Model image of all cluster members before the convolution with the PSF. The image is $\mathrm{2\times2}$ arcmin and covers the entire HST field of view. 
Right: Original, model and residual images of some interacting sources (top,
middle and bottom panel respectively) corresponding to the selection box in the main image ($\mathrm{\sim 9\times 9}$ arcsec). The seven sources 
in the original and residual images are foreground/background galaxies which have not been modeled.}
\label{f:fit}
\end{figure*}

With this procedure, we derive for all cluster members the best-fit values of the parameters of a 
Sersic model (i.e., the effective radius, the magnitude within that radius, the Sersic index, the 
minor to major axis ratio and its position angle). \\
We perform several tests to confirm that our fits are robust. In particular, we check that using 
different PSFs the best-fit values of the parameters obtained for each galaxy are consistent within the errors and that the  global residual image remains unaltered. 
In the left panel of  Figure~\ref{f:fit}, we show the model image of all selected cluster 
members plus the intracluster light in M0416. An example of the goodness of our fits is shown on the right of Figure~\ref{f:fit}.  Panel (a) shows a stamp of the original image containing three 
cluster members and seven foreground/background galaxies. Panel (b) shows the model image of just 
the member galaxies, and Panel (c) shows the residual (i.e., observed minus model) image. \\

\subsection{Stellar mass-to-light ratio}
\label{ss:mtl}
We use the sample of spectroscopically confirmed members for which we have measured the stellar mass 
values (see Sect.~\ref{ss:mass}) to  calibrate the average mass-to-light ratio of all cluster members. From the 62 cluster members with 
stellar mass estimates we exclude three objects which are outside the HFF field of view and two 
objects that show uncommonly high residuals in the fit. The faintest object in this sample has a magnitude of $\mathrm{\sim 23\, mag}$ in the $F160$ band. 
We sum the best-fit stellar mass values of the 57 members and divide this quantity by the total 
luminosity of the same galaxies estimated from our surface brightness modeling. We also employ the 
values of $M_{*}^{low}$ and $M_{*}^{high}$ to derive a minimum and a maximum average stellar 
mass-to-light ratio. To estimate the stellar mass values of all cluster members and thus the stellar 
mass map of the cluster, we use the global model image produced by \galfit before the convolution with the 
PSF and with a zero background and multiply for with the average stellar mass-to-light ratios determined previously.

\section{Baryon to total mass profiles}\label{s:baryon_to_total}
\begin{figure*}[t]
\centering
 \includegraphics[scale=0.8]{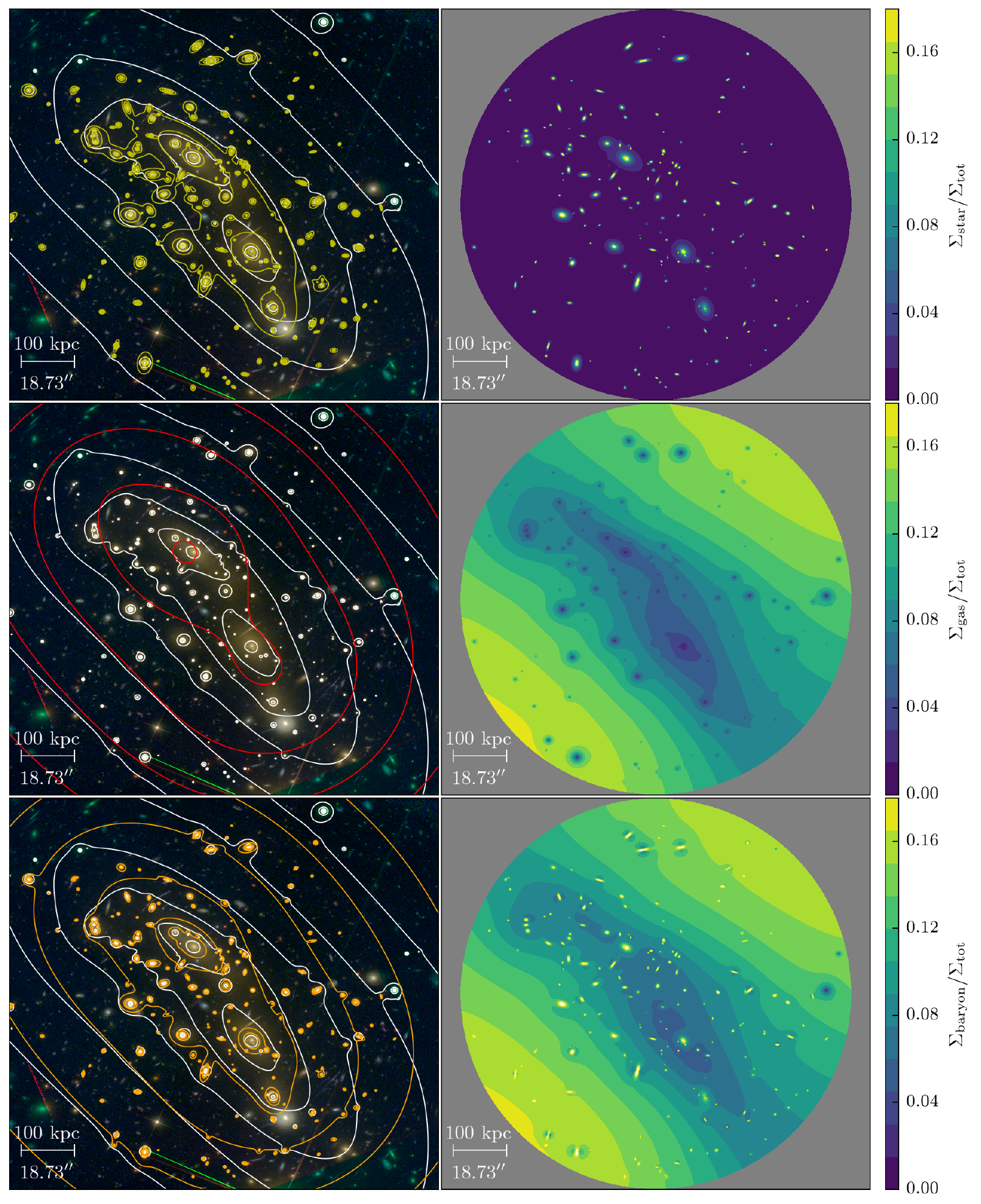}
 \caption{ Left panels: total, stellar, hot gas and baryonic surface mass density isocontours 
overlaid on a color-composite image of the cluster in 7 optical filters.  White lines are the total 
mass isodensity levels which have a logarithmic step between 0.00035 and 0.003 
$\mathrm{\msun/kpc^2}$. Yellow lines refer to the stellar mass isodensities and are spaced between 
$\mathrm{3.5\times 10^{-6}\, to\,  3\times 10^{-4} \msun/kpc^2}$. Red lines are the hot gas isodensity contours spaced between $\mathrm{4.5\times 10^{-5}\, and\, 1.8\times 10^{-4} \msun/kpc^2}$. Orange lines refer to the baryonic component (stars + hot gas) and have the same range as the hot gas component. Right panels show the two-dimensional surface density profile ratios of stellar (upper panel), hot gas (middle panel) and baryon (bottom panel) over total mass.}
 \label{f:sigmatot}
\end{figure*}

\begin{figure*}[t]
\begin{tabular}{ll}
 \includegraphics[width=\columnwidth]{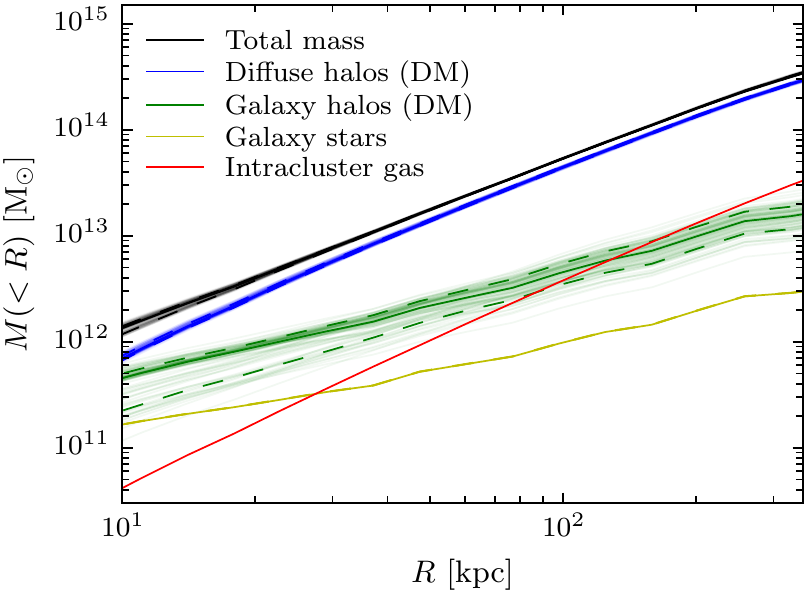} & 
  \includegraphics[width=\columnwidth]{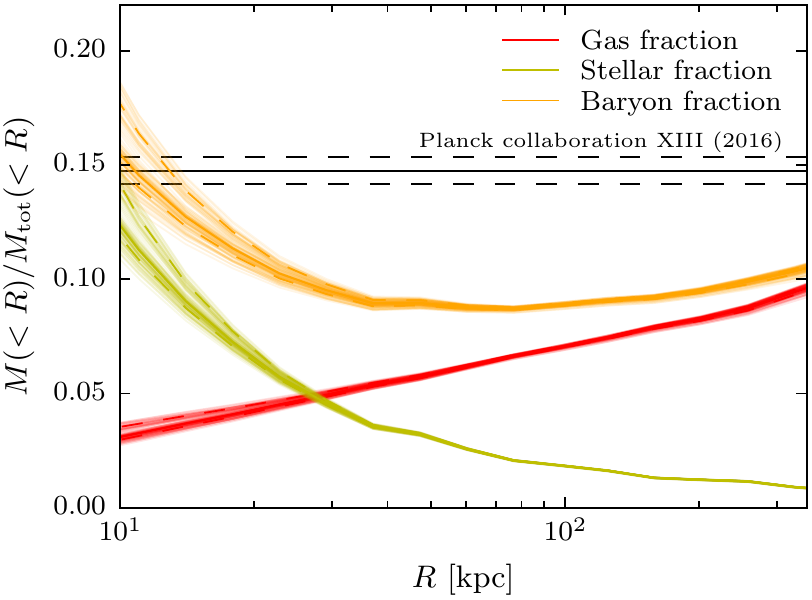} 
\end{tabular}
 \caption{Left Panel: Cumulative projected mass profiles of the different cluster mass components. 
Right panel: ratio between the cumulative projected mass profiles of the baryonic components and 
the cumulative projected total mass profile. The solid and dashed black lines represent the value of the cosmological baryon fraction from \cite{planck2016} with the $1\sigma $ uncertainty.}
  \label{f:mrp}
\end{figure*}

In this section, we analyze the stellar, hot gas and baryon to total mass fraction profiles. The 
baryonic mass profile is defined as the sum of the stellar and hot gas components. 
The hot gas and total mass profiles are measured using the method described in
Bo17 and that we will briefly summarize here.
Deep ($297$~ks) Chandra X-ray observations \citep{ogrean2015} are used to measure the hot gas mass
by fitting, in 2D, the X-ray surface brightness map with dual Pseudo-Isothermal
Elliptical (dPIE) mass distributions.
The best model for the hot gas consists of three spherical dPIE components.
In turn, this is used as a fixed mass component in a standard strong lensing analysis
of the cluster, from which the total, diffuse and galaxy halos masses are
measured.
This method allows for a more self-consistent separation of the dark matter and
hot gas components than a traditional approach.
The CLASH and HFF images are complemented with MUSE data, allowing to boost the
number of spectroscopically-confirmed multiple images to $102$, making this one
of the best dataset available for strong lensing analysis of a galaxy cluster.
These data are used to infer the parameters of the cluster mass model, that
consists of three large-scale halos (diffuse DM), the aforementioned hot gas
component and $193$ cluster members halos that include both the galaxy DM and
stellar mass.\\
In Figure~\ref{f:sigmatot} we show the two-dimensional stellar, hot gas and baryonic mass maps. In the 
left panels of Figure~\ref{f:sigmatot} we plot the total, stellar, hot gas and baryonic 
surface mass density isocontours overlaid on a color-composite image of the cluster in 7 HST 
optical filters. Right panels show the two-dimensional maps of the stellar, hot gas and baryon to total 
mass fractions. We can see  from Figure~\ref{f:sigmatot} that the stellar mass is concentrated 
mainly in the center, which is coincident with the position of the northern BCG, of the cluster and is embedded in the cluster members, while the hot gas contribution 
increases moving towards more external regions. \\
Using the same method as in Bo17, we derive the cumulative 
projected radial profile of the stellar mass component. In the first panel of Figure~\ref{f:mrp}, 
we show the cumulative projected mass profile of the different components: total, diffuse halos 
(mostly DM), galaxy halos, stellar and hot gas. This plot complements Figure 4 in 
Bo17 with the addition of the stellar component. 

The statistical errors on stellar mass profile are derived by considering the minimum 
and maximum stellar mass-to-light ratio values defined in Sect.~\ref{ss:mtl}. 
In the second panel of Figure~\ref{f:mrp} we show the cumulative projected stellar 
and hot gas to total mass profiles of the cluster, obtained from the combination of this 
work and the strong lensing modeling (Bo17). 
In this plot, the statistical errors of the stellar mass component are significantly smaller than those 
of the total mass profile. 
We remark that the stellar mass values derived from a SED fitting depend on the adopted stellar templates and 
IMF. 

The relative contribution of the cluster member subhaloes to the total mass profile decreases moving from the 
cluster center, reaching approximately the same value of the hot gas component at a projected 
distance between 100 and 200 kpc. 
The diffuse DM mass component is the dominant one at all radii. 
The cumulative projected stellar over total mass fraction profile has a decreasing trend, 
with a peak value of $f_{\star} \, \sim$ 15\% near the cluster center and a mean value of 2\% at 
100 kpc from the center. The overall trend is in agreement with that found by \cite{grillo2015}. 

The choice of the stellar IMF can change up to approximately a factor of 2 the estimated 
stellar mass to light ratio. \cite{hoag2016} found for the 
same cluster a mean value of $f_{\star}$ of $\approx$ 0.9\% within a square region of 
side $\sim$ 730 kpc and using a diet-Salpeter IMF. 
\cite{bk2014} found a value of $f_{\star} $ of $\sim$ 1 \% for massive clusters (as massive as 
M0416) at redshift $z=0.3$ using a Chabrier IMF \citep{chabrier2003} at different radial ranges. If 
we consider the conversion factor between the different stellar IMFs, our values are consistent with those obtained in these previous works.
We can also evaluate the 
cumulative projected stellar to total mass fraction in cluster members. This fraction 
reaches a maximum value 
of $\approx 35\%$  near the cluster center and drops to a mean value of 15\% 
at larger clustercentric distances. This fraction is in agreement with that estimated in the 
cores of SDSS massive early-type galaxies \citep[e.g. ][]{grillo2010}. \\
The cumulative projected baryonic to total mass fraction, considered as the summed 
contribution of galaxy stars and hot intracluster gas, starts from 15\% in the cluster core, then 
has a minimum and finally increases up to approximately 10\% at a projected distance of 350 kpc 
from the cluster center.
These trends are in agreement with those found in \cite{bs2006}, who analyzed the mean profiles of 
different mass components by using data from  59 nearby clusters from the ESO Nearby Abell Cluster 
Survey. The value of the baryonic mass fraction at large radii is also comparable with that obtained by 
\citet{gonzales2013} in clusters of similar mass. This fraction is smaller than the 
cosmological baryon fraction estimated from CMB measurements from Planck \citep[$0.147 \,\pm\, 
0.006$,][]{planck2016}. However, this is not very surprising since this analysis is limited to the inner 300 kpc of M0416.

\section{Dark Matter profile}\label{s:dm_profile}

\begin{figure*}[ht]

  \includegraphics[width=\linewidth]{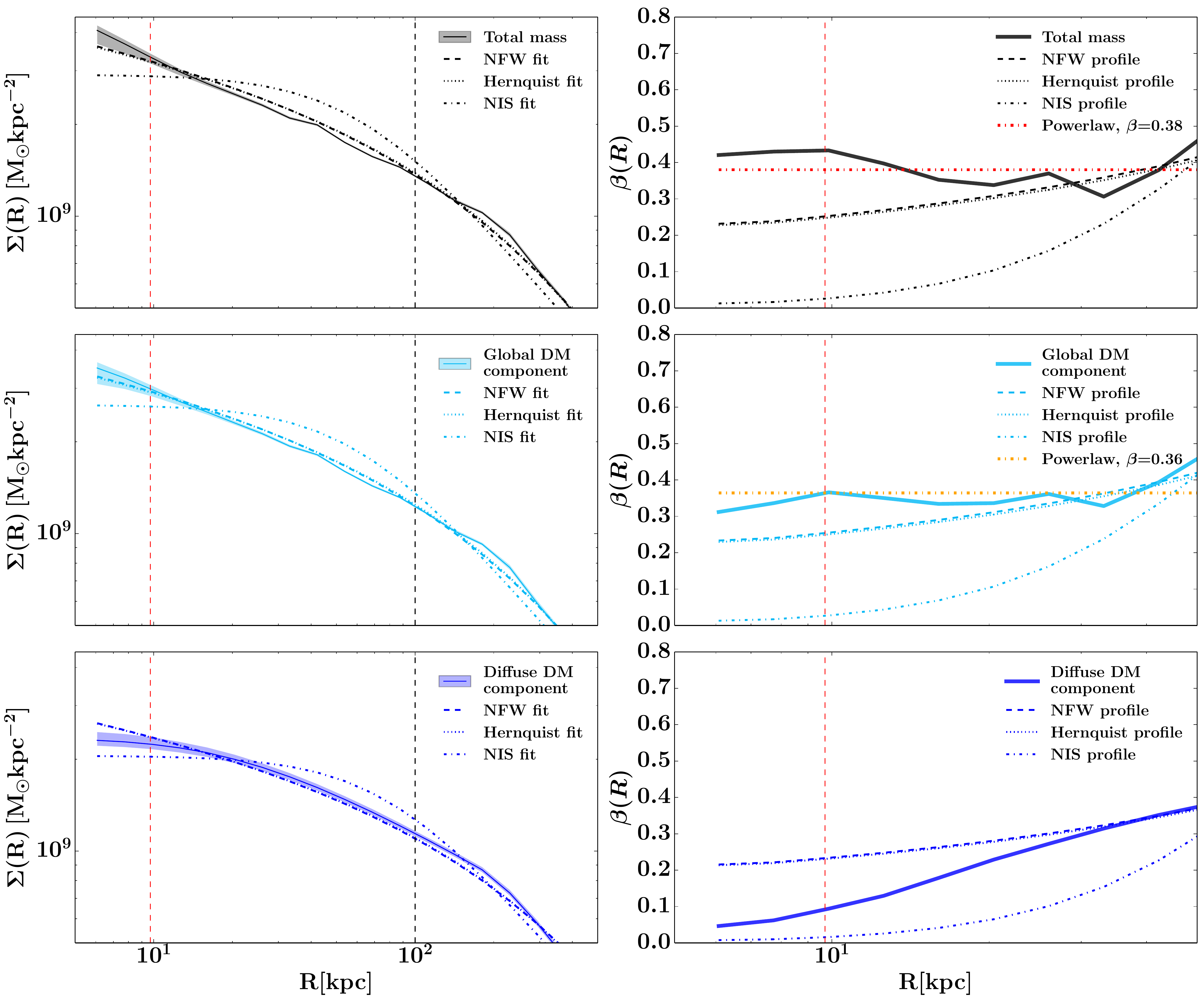}
 \caption{Left Panels: Surface mass density profiles of total (top row) and dark-matter (middle and bottom rows) components fitted with a NFW 
\citep{navarro1996}, Hernquist \citep{hernquist1990} and a softened isothermal sphere \citep[NIS; ][]{grogin1996} profile. 
Right panels: $\beta (R)$ as defined as in Eq.~\ref{e:beta} calculated for the total (top row) and dark matter (middle and bottom rows) surface mass densities and compared with the values of NFW (dashed), Hernquist (dotted), softened isothermal sphere (dot-dashed), fits. The red dashed lines show the effective radius of the 
BCG. The black dashed lines represent the radius where we see a change in the slope of the 
 total surface mass density profile and represents also the range in which we have performed a fit with a  power-law profile. To better show the differences, in the right panels we focus on the inner 50 kpc.}  
  \label{f:nfwfits}
\end{figure*}

In this section we analyze the surface mass density profiles of the different mass components. We 
remark that with our analysis we can disentangle the dark-matter only component, as opposed to 
most of the previous studies.\\
In the left panels of Figure ~\ref{f:nfwfits}, we show the total matter, diffuse and global 
dark-matter surface mass density profiles of M0416 fitted with NFW, \cite{hernquist1990} 
softened isothermal sphere \citep[NIS; ][]{grogin1996} and power-law profiles. The global dark-matter 
component is the sum of the diffuse term and that embedded in galaxy halos. The latter has been 
obtained from the total mass density profiles of the galaxies reconstructed in the lensing 
optimization and subtracting their stellar mass density profiles described above.
From this plot, we can see that the total and global dark-matter surface mass density profiles are overall 
well fitted by NFW, Hernquist, and in the inner 100 kpc, power-law  profiles, while a NIS profile provides a poor fit.  For this reason, a NIS profile is not considered in the following.\\ 
We compare the values of the parameters of the best-fitting NFW profile we derive for the 
total mass with those obtained from the weak lensing analysis by \cite{umetsu2014}. Our estimate of 
$M_{200, c}$ is  $1.6 \times 10^{15} \msun$,  \cite{umetsu2014} measured $(1.04 \pm 
0.22)\times 10^{15}  \, \msun$ for the same cluster but with a slightly different cosmology. 
The discrepancy between these values might suggest that the fits of the separate strong and weak lensing data
cannot be used to extrapolate correctly in the outer and inner regions, respectively, the total mass of M0416.
However, we remark that the independent strong and weak lensing total mass estimates of M0416 nicely match in the overlapping region (see Fig. 16 in \citealt{grillo2015}) and are overall consistent with the results from the X-ray and dynamical analyses (see Fig. 13 in \citealt{balestra2016}).

From the left panels of Figure ~\ref{f:nfwfits}, we infer that due to projection effects, for models 
with 
two different inner and outer slopes, the sum of multiple components (M0416 does not have a unimodal 
total mass distribution), and the ``circularization'' ($\mathrm{M(<R)}$ and $\mathrm{\Sigma(R)}$) of the profiles, 
contribute to the result of obtaining more than one model that is consistent with the reconstructed 
surface mass density profiles.
In the very central region ($\mathrm{R \, \lesssim 15 kpc}$), both the total matter and the global dark-matter profiles are 
steeper than cored profiles and NFW profiles. On the contrary, the DM 
diffuse profile is flatter. This can be explained considering that the center of the cluster is 
coincident with the position of the northern BCG, hence the steep total and global dark-matter 
profiles can be related to the dark-matter halo of the BCG. \\
In the right panels of Figure~\ref{f:nfwfits}, we show the radial dependence of the logarithmic slope 
$\beta (R)$, defined as
\begin{equation}
\beta (R) = - \mathrm{\frac{dln\Sigma (R)}{dln(R)}}. 
\label{e:beta}
\end{equation}
In the case of a power-law profile  $\mathrm{\gamma_{in} = \beta\, +\, 1}$ (see Sect.~\ref{s:intro}). In the other cases the relation between the two slopes is more complex.
This quantity has been calculated numerically for the total matter and global dark-matter surface profiles, as derived from the data and as predicted from the NFW, Hernquist, NIS, and power-law best-fits. We evaluate the slope values only in the radial range between 5 and 100 kpc.
In the very center of the cluster (i.e., the region dominated by the northern BCG), the 
total matter and global dark-matter density profiles are steeper than NFW and Hernquist models. The global
DM profile is somewhat flatter than the total mass profile. The diffuse 
DM profile is significantly flatter than the others. In the diffuse DM component the BCG is not included. 
The best-fit values of the  logarithmic slope of the power-law fits, within the inner 100 kpc, are $\beta_{tot} \, =\, 0.38\, \pm 
0.01$ and $\beta_{DM}\, =\, 0.36\, \pm 0.01$ for the total matter
and global dark-matter profiles, respectively. We do not fit with this model the diffuse DM 
component because the slope cannot be well approximated by a single value.\\
Comparing our results to previous works is difficult because so far most observational studies have only focused on the total mass density profile (see Sect.~\ref{s:intro}). 
\citet{newmana,newmanb} use a small sample of massive ($\mathrm{M_{200} = 0.4}$ - $\mathrm{2 \times 
10^{15} \msun}$), relaxed galaxy clusters, at $\mathrm{z = 0.19-0.31}$, to measure the  DM 
inner ($\mathrm{\lesssim 30 kpc}$) slope and compare it with that of the total 
mass and that predicted by simulations for collisionless dark-matter halos. They found that the slope of the observed total mass density profile $\mathbf{(<\gamma_{in}>\, = 1.16 \pm\,  0.05 )}$ is in agreement with that predicted  from DM only simulations. They proposed a scenario according to which an early dissipative phase of star formation in the BCG establishes a 
steeper total mass density profile ($\mathrm{\rho_{tot}}$) in the inner regions of a cluster (5 $-$ 10 kpc). The subsequent accretion of stars then mostly replaces the dark-matter, so that the total mass profile is nearly maintained. 
In the same works, the observed DM profile was found to be significantly shallower $\mathbf{(<\gamma_{in}>\, = 0.5 \pm\,  0.1 )}$ than canonical NFW models in the radial  range  $r \mathrm{\, \lesssim \, 30 kpc}$, comparable with the effective radius of the BCG. 
 In \citet{newmanb} it is argued that variations  in  the observed inner 
 dark-matter profiles can be seen from cluster to cluster, correlating with the
size and mass of the BCG.  This would  suggest a connection between the dark-matter profile in the
cluster cores and the assembly of stars in the BCGs. \citet{laporte2014},  using state of the art $N-$body 
resimulations of the growth of rich galaxy clusters between z=2 and z=0, show that the steeper and shallower
profiles of, respectively,  total mass and global dark-matter found by \citet{newmanb} can be explained as the result of dissipationless mergers.
Note that \citet{newmanb} and 
\citet{laporte2014} adopt definitions of the dark-matter profiles slightly different from ours. In \citet{newmanb} the dark-matter mass density profile was obtained by subtracting from the total mass density profile that of the BCG stellar component. Further, \citet{newmanb} did not 
consider the intracluster hot gas, claiming that its inclusion would not change the shape of 
the dark-matter density profile. The definition of dark-matter is close to
our definition of global dark-matter. The 
dark-matter component of \citet{laporte2014} is comparable to our definition of global 
dark-matter component (by construction). With this in mind, we do find an 
indication that the global dark-matter is flatter than the total component, even if this difference is 
not significant. On the other hand, from Figure 5 of  \citet{newmanb} a cluster with a BCG as small as the one in M0416 
(in  terms of effective radius)  is expected to have a steeper dark-matter profile, 
hence much more similar to the total one. 

\vspace{2cm}

\section{Conclusions}\label{s:conclusion}
In this work, we have decomposed the total mass profile of the galaxy cluster 
MACS J0416$-$2403 into its different components: stellar, hot gas, dark-matter diffuse and dark-matter substructures. To this aim, we have used state of the art lens models based on HFF imaging data and extensive VLT spectroscopy, as well as deep Chandra observations. We have determined the cumulative projected radial mass profiles and the surface mass density maps of these components.
 For the first time, we have been able to separate all components with little previous assumptions and also to map precisely the dark-matter only distribution within 300 kpc from the cluster center.  \\
Our main results can be summarized as follows.
\begin{itemize}
\item The stellar and hot gas components are only a small percentage of the total matter 
in the cluster. The stellar mass contribution reaches the peak value of $f_*\, =\, 15\%$ within 20 
kpc from the cluster center, due to the presence of the BCG, then decreases to a mean value of 2\% 
at 100 kpc from the cluster center. The hot gas to total mass fraction, instead, increases with the distance from 
the center. The baryon fraction, evaluated as the sum of the stellar and hot gas components over the total mass of  the cluster, has a peak value of 15\% in the cluster center, then decreases reaching $\sim$ 10\% at 350 kpc. Both the stellar and baryon fractions are in general good agreement with the global values  found in the literature. Our baryon fraction is smaller than the cosmological baryon fraction measured by \citet{planck2016}, which however refers to large radii.
\item We have evaluated the ratio between the stellar and total mass embedded in substructures. This 
fraction is $\sim$ 30\% near the cluster center, then decreases to $\mathrm{\sim}$ 15\% at larger clustercentric 
distances. 
\item We have studied the total mass, global and diffuse dark-matter surface density profiles.
In the radial range between 5 and 50 kpc, the surface mass density profiles of the total mass and
global dark-matter have comparable slopes. In this radial range, if we parametrize
$\Sigma (R)$ as $R^{-\beta}$, we obtain values of $\mathrm{\beta}$ equal to $0.38\, \pm\, 0.01$ and $ 0.36\, \pm\,  0.01$ for the total and global dark-matter, respectively. These profiles appear steeper than a NFW profile. The diffuse dark-matter component has a profile much flatter near the cluster center 
that cannot be approximated with a power-law. The 
difference among these three profiles is related to the BCG dark-matter halo and persists  up to $\mathrm{\sim 30\, kpc}$  from the cluster center,  which is approximately three times the value of the effective radius of the BCG.  
\item As a result of to the mass decomposition presented in this work, we are able to
confirm previous findings from \cite{carminha2017} and Bo17 regarding the absence
of a significant ($>3 \sigma$) offset between the dark-matter and the stellar (BCGs) components.
A secure detection of such offsets in merging systems would be important, since they are predicted by models of self-interacting dark matter \citep[e.g.][]{markevitch2006}. We remark, however, that despite the accurate modeling of DM and baryonic components developed here, it remains very difficult to establish the presence of offsets of a few arcsec, due to a number of inherent systematics in the lens model, as well as line-of-sight lensing effects \citep{chirivi2017}. 
\end{itemize}

\noindent In the future, we plan to extend this analysis to other clusters from the CLASH sample with highly precise strong lensing data and MUSE spectroscopy.

\section*{Acknowledgements}
We acknowledge support from PRIN-INAF 2014 1.05.01.94.02 (PI M. Nonino). M.B. and C.G. acknowledge support by the VILLUM FONDEN Young Investigator Programme through grant no. 10123.

\bibliographystyle{aasjournal}
\bibliography{bibliography}

\end{document}